\documentclass[aps,showpacs,nofootinbib,twocolumn]{revtex4}
\usepackage[dvips]{graphicx}
\usepackage{amssymb}
\newcommand{\ped}[1]{\ensuremath{_{\rm #1}}}
\newcommand{\apex}[1]{\ensuremath{^{\rm #1}}}

\begin{document}

\title{Effects of paramagnetic impurities on two-band superconductor}

\author{G.A. Ummarino}\email{E-mail:giovanni.ummarino@infm.polito.it}

\affiliation{ INFM-Dipartimento di Fisica, Politecnico di Torino,
Corso Duca degli Abruzzi 24, 10129 Torino, Italy}

\begin{abstract}
I calculated the effect of magnetic impurities on the normal and
superconductive properties of a multiband s-wave superconductor by
direct solution of the two-band Eliashberg equations. In this way I
determined the critical temperature, the values of the
superconductive gaps, the shape of the superconductive density of
states and other physical quantities that depend on the
concentration of magnetic impurities. I found that the gaps and the
penetration lengths display an unusual behaviour as a function of
temperature. I examine the possibility that the presence of a
negative induced gap raises the critical temperature.
\end{abstract}


\maketitle
A specific, but representative case of anisotropic superconductivity
is multiband superconductivity where the order parameter is
different in different bands. The more famous and clear case is the
magnesium diboride \cite{Nagamatsu} where the two gaps are observed
in several experiments. Some properties of this material changes
markedly from the BCS s-wave one band case. For example the rate
between the gaps and the critical temperature can be major or minor
then 3.53. Another difference is that the non magnetic interband
scattering rate produces a reduction of the critical temperature
\cite{carbinicol} while in an one band superconductor there isn't
effect on the critical temperature. Recently also the effects of
magnetic impurities on a multiband superconductor have been studied,
in a theoretical \cite{moca,nostroPRL} and experimental way, on MgB2
doped by Mn \cite{nostroPRL,Rogacki}.

Let us start from the generalization of the Eliashberg theory
\cite{Eliashberg} for systems with two bands \cite{Kresin,Golimp}
,that has already been used with success to study the MgB2 system
\cite{Brinkman,Choi,GolubovA2F,Johntony}. To obtain the gaps and the
critical temperature within the s-wave, two-band Eliashberg model
one has to solve four coupled integral equations for the gaps
$\Delta_{i}(i\omega_{n})$ and the renormalization functions
$Z_{i}(i\omega_{n})$, where i is a band index and $\omega_{n}$ are
the Matsubara frequencies. I included in the equations the
non-magnetic and magnetic impurity scattering rates in the Born
approximation, $\Gamma^{N}_{ij}$ and $\Gamma^{M}_{ij}$.
 \begin{eqnarray}
\omega_{n}Z_{i}(i\omega_{n})&=&\omega_{n}+\pi
T\sum_{m,j}\Lambda_{ij}(i\omega_{n}-i\omega_{m})N^{j}_{Z}(i\omega_{m})+\nonumber\\
& &
+\sum_{j}(\Gamma^{N}_{ij}+\Gamma^{M}_{ij})N^{j}_{Z}(i\omega_{n})\label{eq:EE1}
\end{eqnarray}
\begin{eqnarray}
Z_{i}(i\omega_{n})\Delta_{i}(i\omega_{n})&=&\pi
T\sum_{m,j}[\Lambda_{ij}(i\omega_{n}-i\omega_{m})-\mu^{*}_{ij}(\omega_{c})]\cdot\nonumber\\
& &
\hspace{-2.5cm}\cdot\theta(|\omega_{c}|-\omega_{m})N^{j}_{\Delta}(i\omega_{m})+\sum_{j}%
(\Gamma^{N}_{ij}-\Gamma^{M}_{ij})N^{j}_{\Delta}(i\omega_{n})
\label{eq:EE2}
\end{eqnarray}
 where $\theta$ is the Heaviside function, $\omega_{c}$ is a
 cut-off energy and
 $\Lambda_{ij}(i\omega_{n}-i\omega_{m})=\int_{0}^{+\infty}d\omega
\alpha^{2}_{ij}F(\omega)/[(\omega_{n}-\omega_{m})^{2}+\omega^{2}]$,
$N^{j}_{\Delta}(i\omega_{m})=\Delta_{j}(i\omega_{m})/
{\sqrt{\omega^{2}_{m}+\Delta^{2}_{j}(i\omega_{m})}}$,
$N^{j}_{Z}(i\omega_{m})=\omega_{m}/
{\sqrt{\omega^{2}_{m}+\Delta^{2}_{j}(i\omega_{m})}}$.

The solution of the Eliashberg equations requires as input: i) the
four (but only three independent \cite{Kresin}) electron-phonon
spectral functions $\alpha^{2}_{ij}(\omega)F(\omega)$; ii) the four
(but only three independent \cite{Kresin}) elements of the Coulomb
pseudopotential matrix $\mu^{*}(\omega\ped{c})$; iii) the two (but
only one effective \cite{Kresin}) non-magnetic impurity scattering
rates $\Gamma^{N}_{ij}$; iv) the four (but only three independent
\cite{Kresin}) paramagnetic impurity scattering rates
$\Gamma^{M}_{ij}$.  The four spectral functions
$\alpha^{2}_{ij}(\omega)F(\omega)$, that were calculated for pure
$MgB_{2}$ in ref. \onlinecite{GolubovA2F}, have the following
electron-phonon coupling constant:
$\lambda_{\sigma\sigma}(x\!\!=\!\!0)$=1.017,
$\lambda_{\pi\pi}(x\!\!=\!\!0)$=0.448,
$\lambda_{\sigma\pi}(x\!\!=\!\!0)$=0.213 and
$\lambda_{\pi\sigma}(x\!\!=\!\!0)$=0.156.
%

As far as the Coulomb pseudopotential is concerned, I use the
expression calculated for pure MgB$_2$ \cite{DolgovCoulomb}:
\begin{eqnarray}
\hspace{-5mm}\mu^{*}\!\!&\!=\!&\!\! \left| \begin{array}{cc}%
\mu^{*}\ped{\sigma \sigma} & \mu^{*}\ped{\sigma \pi}\\
\mu^{*}\ped{\pi \sigma} & \mu^{*}\ped{\pi \pi}
\end{array} \right| =  \nonumber \\
\!\!&\!=\!&\!\! \mu(\omega_{c})N\ped{N}\apex{tot}(E\ped{F})
\left| \begin{array}{cc}%
\frac{2.23}{N\ped{N}\apex{\sigma}(E\ped{F})} &
\frac{1}{N\ped{N}\apex{\sigma}(E\ped{F})}\\ & \\
\frac{1}{N\ped{N}\apex{\pi}(E\ped{F})} &
\frac{2.48}{N\ped{N}\apex{\pi}(E\ped{F})}
\end{array} \right| \label{eq:mu}
\end{eqnarray}
\begin{figure}[t]
\includegraphics[keepaspectratio,width=0.65\columnwidth,angle=270]{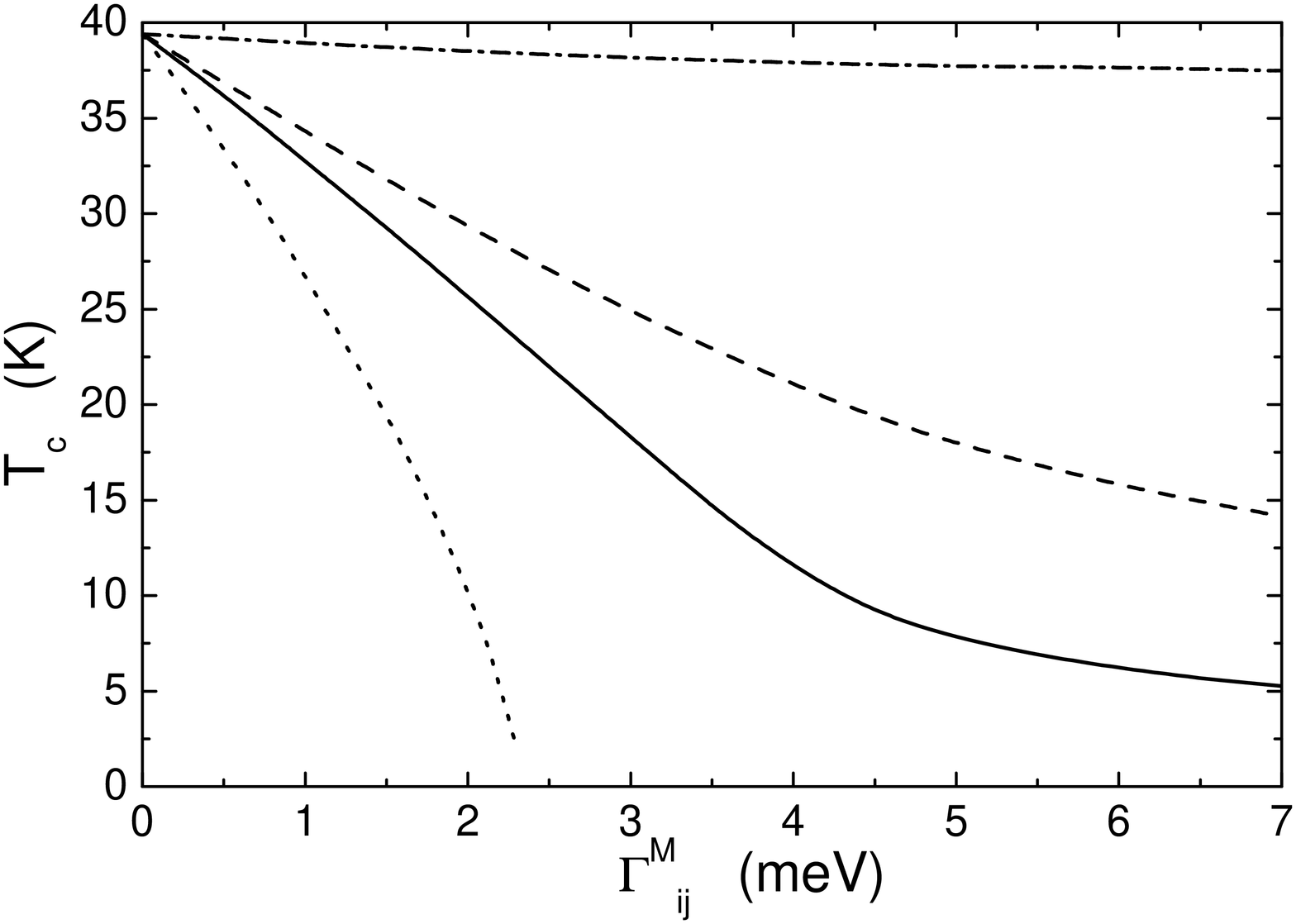}
\vspace{-0mm} \caption{The calculated critical temperature as
function of $\Gamma\apex{ij}_{-}$:
$\Gamma\apex{\sigma\sigma}_{-}\neq0$ (solid line),
$\Gamma\apex{\sigma\pi}_{-}\neq0$ (dashed line),
$\Gamma\apex{\sigma\sigma}_{-}=\Gamma\apex{\sigma\pi}_{-}\neq0$
(dotted line) and $\Gamma\apex{\pi\pi}_{-}\neq0$ (dashed-dotted
line).} \label{fig:Fig1S06}
\end{figure}
\begin{figure}[!]
\begin{center}
\includegraphics[keepaspectratio,width=\columnwidth]{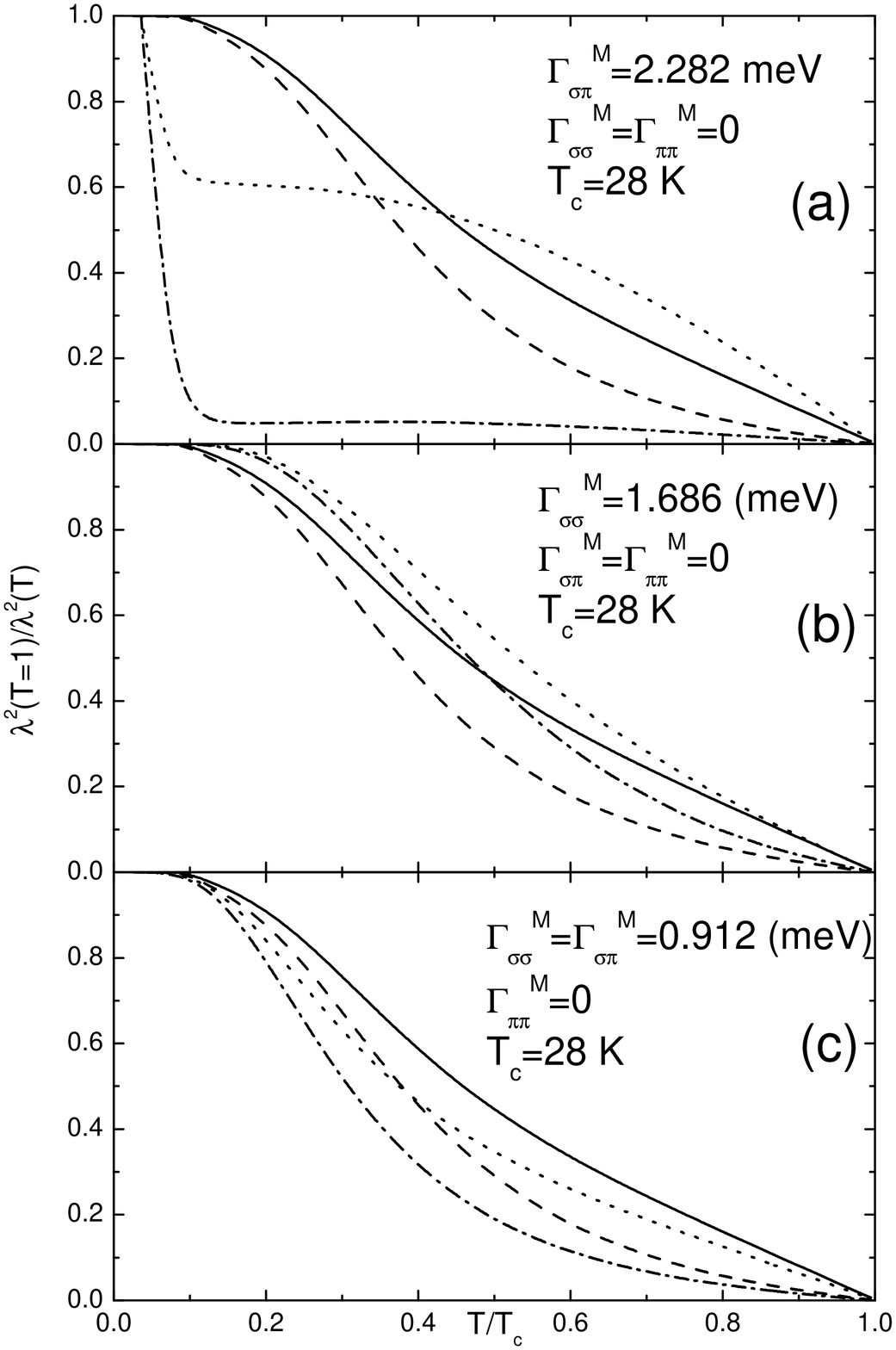}
\end{center}
\vspace{-5mm} \caption{The calculated penetration depth as function
of temperature: panel a: pure $MgB_{2}$ case (solid line ab plane,
dashed line c axis), $\Gamma\apex{\sigma\sigma}_{-}\neq0$ (dotted
line ab plane, dotted-dashed c axis); panel b: pure $MgB_{2}$ case
(solid line ab plane, dashed line c axis),
$\Gamma\apex{\sigma\sigma}_{-}=\Gamma\apex{\sigma\pi}_{-}\neq0$
(dotted line ab plane, dotted-dashed c axis);panel c: pure $MgB_{2}$
case (solid line ab plane, dashed line c axis),
$\Gamma\apex{\sigma\pi}_{-}\neq0$ (dotted line ab plane,
dotted-dashed c axis).} \label{fig:Fig2S06}
\end{figure}
where $\mu(\omega\ped{c},x)$ is a free parameter and
$N\ped{N}\apex{tot}(E\ped{F},x)$ is the total normal density of
states at the Fermi level. For obtaining the experimental critical
temperature of pure $MgB_{2}$ case ($T_{c}=39.4$ K) we fix
$\mu(\omega_{c})=0.03105$ with cut-off energy $\omega_{c}=500$ meV
and maximum energy 570 meV. In all our calculations I use
\cite{GolubovA2F} $N\ped{N}\apex{\sigma}(E\ped{F})=0.30$
states/(cell eV) and $N\ped{N}\apex{\pi}(E\ped{F})=0.41$
states/(cell eV). In this work I study only the effect of
paramagnetic impurities so is always $\Gamma^{N}_{ij}=0$. In fig. 1
we can see the effect of magnetic impurities on the critical
temperature in four limit cases. The stronger reduction of $T_{c}$
is when the $\Gamma^{M}_{\sigma\sigma}$ is preponderant while the
effect of the $\Gamma^{M}_{\pi\pi}$ is week. A part the
$\Gamma^{M}_{\pi\pi}$ cases a small amount of impurities is safe for
reduce $T_{c}$ in a considerable way. This is consistent also with
the use of the Born approximation and with the fact of thinking the
$MgB_{2}$ doped with paramagnetic impurities as a system perturbed
and not as new material with different electron-phonon coupling
constant, phonon energies and Coulomb pseudopotentials.
In fig. 2 we can see the calculated penetration depth by solution of
imaginary axis Eliashberg equations \cite{Dolghilambda}. The curve
are remarkably different of pure case especially when is present
interband scattering.

In fig. 3 panel a we can see the calculated values of
$\Delta_{i}(i\omega_{n=0})$ as function of the temperature always
with $T_{c}=28K$. When are present magnetic impurities in the
interband channel the curve has a maximum more evident if the
impurity content is bigger. In the $\Gamma^{M}_{\sigma\pi}\neq0$
case the $\Delta_{\pi}(i\omega_{n=0})$ gap is negative in a small
range of temperature (experimentally it is found that the presence
of magnetic interband scattering is very small \cite{nostroPRL}). In
the panel b we can see the calculated tunneling conductance at $T=4
K$: in this case the curves are similar for the effect of smearing
of the temperature: it is necessary to go to lower temperatures for
noting appreciable differences. Of course from experiments of
quasiparticle tunneling it is impossible to determine the sign of
$\Delta_{\pi}$. In the fig. 4 it is possible see the calculated
values of $\Delta_{i}(i\omega_{n=0})$ as function of the temperature
with $T_{c}=18K$ (panel a) and $T_{c}=13.3K$ (panel b). In the panel
b we can note that, in the $\Gamma^{M}_{\sigma\sigma}\neq0$ case,
there is a range of temperatures where
$\Delta_{\pi}(i\omega_{n=0})>\Delta_{\sigma}(i\omega_{n=0})$.

\begin{figure}[!]
\begin{center}
\includegraphics[keepaspectratio,width=\columnwidth]{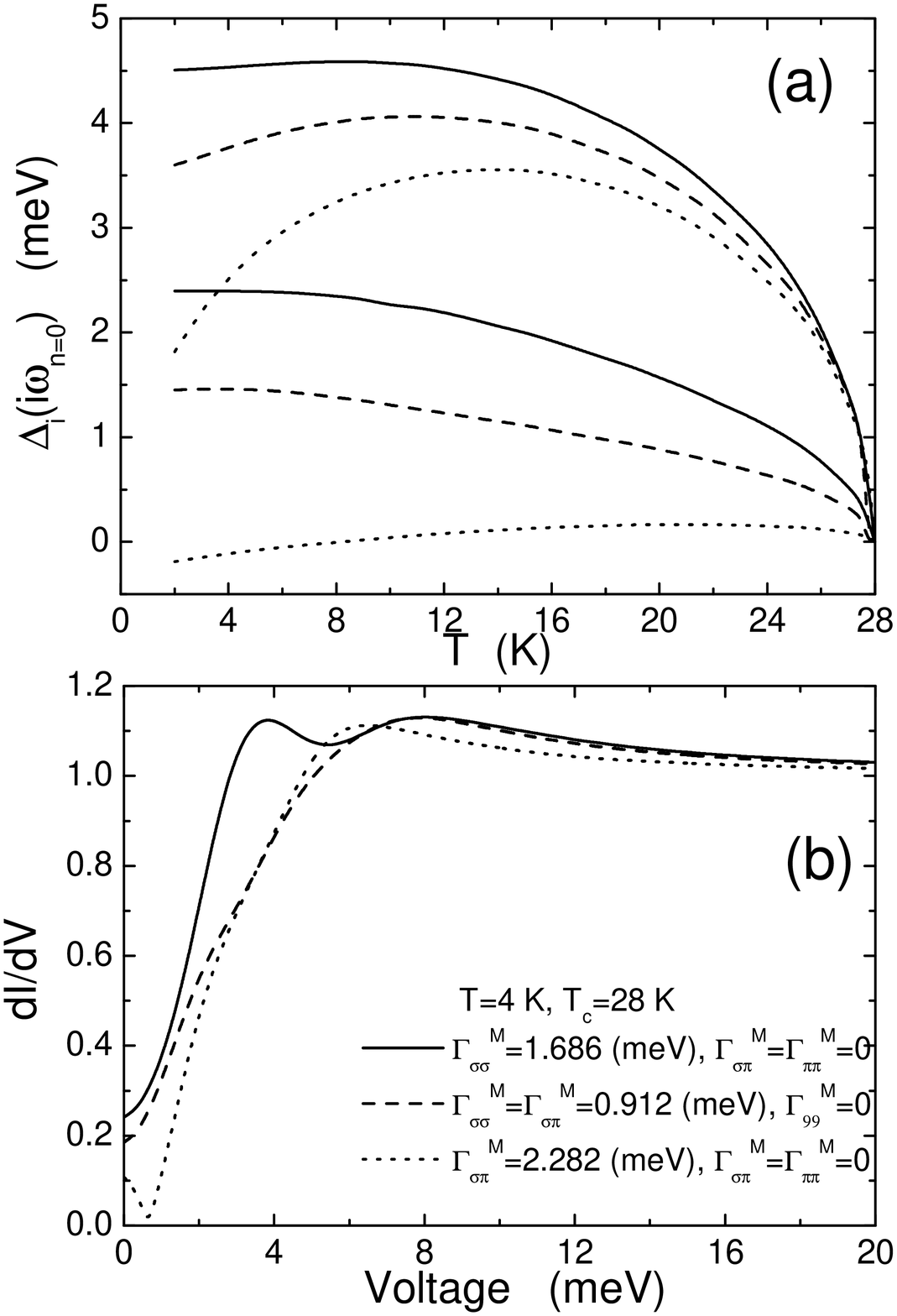}
\end{center}
\vspace{-5mm} \caption{Panel a: the calculated
$\Delta_{i}(i\omega_{n=0})$, when $T_{c}=28K$, as function of
temperature: $\Gamma\apex{\sigma\sigma}_{-}\neq0$ (solid line),
$\Gamma\apex{\sigma\sigma}_{-}=\Gamma\apex{\sigma\pi}_{-}\neq0$
(dashed line), $\Gamma\apex{\sigma\pi}_{-}\neq0$ (dotted line);
panel b: the calculated tunneling conductance, at $T=4K$, in the
three different magnetic doping case:
$\Gamma\apex{\sigma\sigma}_{-}\neq0$ (solid
line),$\Gamma\apex{\sigma\sigma}_{-}=\Gamma\apex{\sigma\pi}_{-}\neq0$
(dashed line), $\Gamma\apex{\sigma\pi}_{-}\neq0$ (dotted line)}
\label{fig:Fig3S06}
\end{figure}
\begin{figure}[!]
\begin{center}
\includegraphics[keepaspectratio,width=\columnwidth]{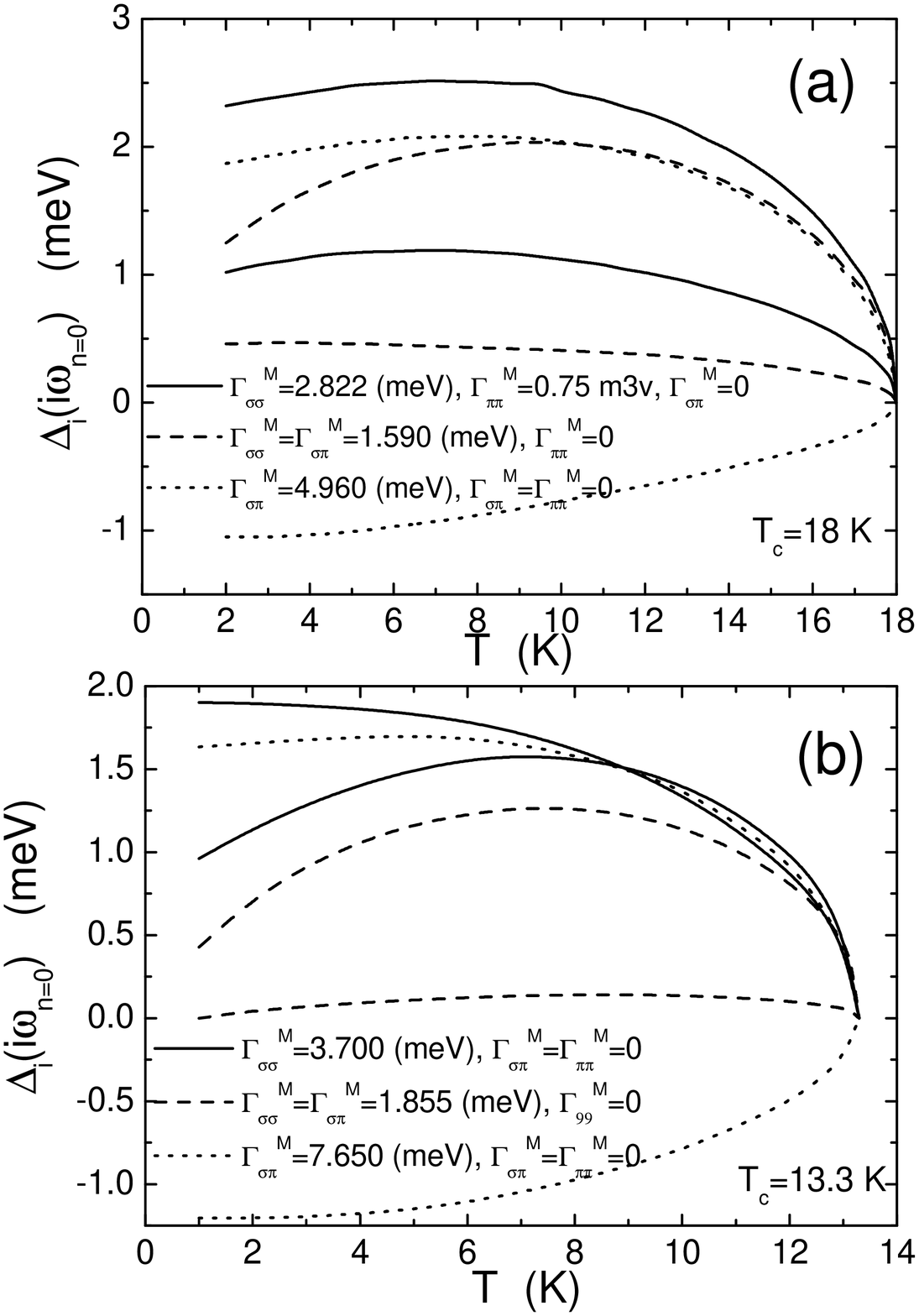}
\end{center}
\vspace{-5mm} \caption{Panel a: the calculated
$\Delta_{i}(i\omega_{n=0})$, when $T_{c}=18K$, as function of
temperature: $\Gamma\apex{\sigma\sigma}_{-}\neq0$ (solid line),
$\Gamma\apex{\sigma\sigma}_{-}=\Gamma\apex{\sigma\pi}_{-}\neq0$
(dashed line), $\Gamma\apex{\sigma\pi}_{-}\neq0$ (dotted line);
panel b: the calculated $\Delta_{i}(i\omega_{n=0})$, when
$T_{c}=13.3K$, as function of temperature:
$\Gamma\apex{\sigma\sigma}_{-}\neq0$ (solid line),
$\Gamma\apex{\sigma\sigma}_{-}=\Gamma\apex{\sigma\pi}_{-}\neq0$
(dashed line), $\Gamma\apex{\sigma\pi}_{-}\neq0$ (dotted line)).}
\label{fig:Fig4S06}
\end{figure}
\begin{figure}[!]
\includegraphics[keepaspectratio,width=0.65\columnwidth,angle=270]{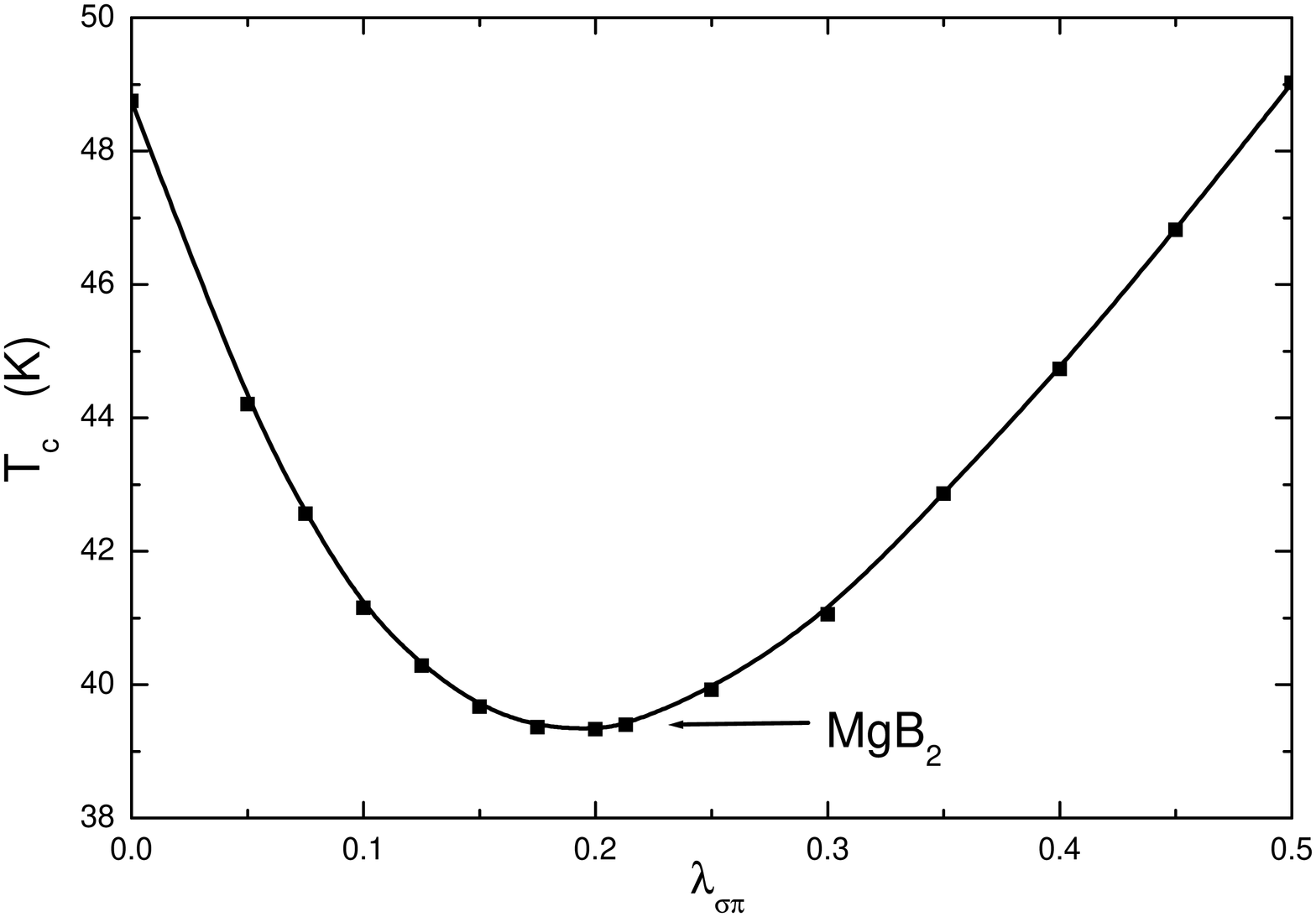}
\vspace{-0mm} \caption{The theoretical critical temperature of an
hypothetical superconductor with all parameters less
$\lambda_{\sigma\pi}$ equals to $MgB_{2}$ case.} \label{fig:Fig5S06}
\end{figure}

It is important remember that in presence of strong coupling
electron-phonon interaction or magnetic impurities the value of
$\Delta_{i}(i\omega_{n=0})$ obtained by solving the imaginary axis
Eliashberg equations and the $\Delta_{i}$ obtained by real axis
formulation can be very different so it is necessary always or the
analitical continuation \cite{analiticalcont} or the real axis
solution for determine the physical quantities  comparable to
measurable experimental gaps.

If we put the interband electron phonon coupling constant
$\lambda_{\sigma\pi}$ (of course also $\lambda_{\pi\sigma}$) and
Coulomb pseudopotential matrix $\mu^{*}_{ij}$ equal to zero, in the
$MgB_{2}$ case, I find an hypothetic material with $T_{c}=45.55$ K
and only a superconductive gap in the $\sigma$ band because the
$\pi$ band is superconductive only for the presence of interband
term. In the $MgB_{2}$ case, the presence of interband term reduces
$T_{c}$. If I calculate the critical temperature in function of
$\lambda_{ij}$ with all other parameters equal to $MgB_{2}$ case, I
found that when $\lambda_{\sigma\pi}=\lambda^{MgB_{2}}_{\sigma\pi}$
the $T_{c}$ is next to a minimum as we can see in fig. 5.

Now, if is only  $\lambda_{ij}=0$, $i\neq j$ but the matrix
$\mu^{*}_{ij}$ has the usual $MgB_{2}$ values it can induce a
negative gap in the $\pi$ band and overall raise the critical
temperature until $48.75K$. We can think of obtaining a negative
$\Delta_{\pi}(i\omega_{n=0})$ gap i.e., roughly, when applies the
condition $\lambda_{ij}-\mu^{*}_{ij}<0$, $i\neq j$, with chemical or
field effect doping where x is the doping content. I assume, for
simplicity,
$\lambda_{\pi\sigma}(x)=\lambda_{\pi\sigma}(0)N\ped{N}\apex{\sigma}(E\ped{F},x)/N\ped{N}\apex{\sigma}(E\ped{F},0)$
\cite{Johntony} and $\mu^{*}_{\pi\sigma}(x)=\mu^{*}_{\pi\sigma}(0)
\cdot N\ped{N}\apex{tot}(E\ped{F},x)\cdot
N\ped{N}\apex{\pi}(E\ped{F},0)/(N\ped{N}\apex{\pi}(E\ped{F},x)\cdot
N\ped{N}\apex{tot}(E\ped{F},0))$ so I find, in the $MgB_{2}$ case,
$N\ped{N}\apex{\pi}(E\ped{F},x)<0.0311
N\ped{N}\apex{\sigma}(E\ped{F},x)/(0.3805
N\ped{N}\apex{\sigma}(E\ped{F},x)-0.0311)$. For example in the case
of Al doping, where, of course, I neglect the presence of disorder,
i.e. I put $\Gamma^{N}_{ij}= 0$, this condition is safe only for
$x>0.46$ when the material isn't  more a superconductor! To raise
the critical temperature is more difficult because $T_{c}$ decreases
principally if $\lambda_{\sigma\sigma}$ lowers and, in our rough
model,
$\lambda_{\sigma\sigma}(x)=\lambda_{\sigma\sigma}(0)N\ped{N}\apex{\sigma}(E\ped{F},x)/N\ped{N}\apex{\sigma}(E\ped{F},0)$
decreases with $N\ped{N}\apex{\sigma}(E\ped{F},x)$ so it is
necessary to study in depth the problem. An other way for having a
negative gap is if roughly
$\lambda_{ij}-\mu^{*}_{ij}-\Gamma^{M}_{ij}<0$, $i\neq j$, but the
critical temperature, in this case, decreases: i.e. it is possible
to have an induced negative gap or with interband magnetic
impurities or when the interband coupling is negative for example
without phononic components but with the interband Coulomb
pseudopotential different from zero. The difference in the two cases
is in the structure of Eliashberg equations: the Coulomb term is
present only in the equation of the order parameter (eq. 2) and a
negative gap in the channel $\pi$ produces a positive contribute in
the more important channel $\sigma$ so the critical temperature
raises while in the case of interband impurities in the equation of
the order parameter there is a positive contribute  of  the Coulomb
term but a negative contribute of the impurity term and also, in the
equation of renormalization fuction (eq. 1), a positive contribute
of the impurity term that raises the value of the renormalization
function but, for consequence, lowers the critical temperature.

I want remember that the possibility of a negative induced gap in a
multiband system have been already  previewed by A.A. Golubov et al.
more ten years ago (see ref. 8).

In conclusion the effects of magnetic impurities can produce, in
multiband superconductors, unusual behaviour as the temperature
dependence of the order parameter and penetration depth but overall
can induce superconductivity with negative order parameter in the
eventual normal band.

%
%

\end{document}